\documentclass[twocolumn,secnumarabic,amssymb,nobibnotes,aps,prd,superscriptaddress,nofootinbib]{revtex4-2}

\usepackage[utf8]{inputenc}
\usepackage{graphicx,amssymb,amsmath,amsthm,amsfonts,epsfig,epsf,environ,float,upgreek}
\usepackage[usenames,dvipsnames,svgnames,table]{xcolor}
\usepackage{epstopdf}
\definecolor{darkred}{rgb}{0.5,0,0}
\usepackage{bm}
\usepackage{dcolumn}
\usepackage{latexsym}
\usepackage{tabularx}
\usepackage{rotating}
\usepackage{longtable}

\usepackage{enumerate}
\usepackage{tensor,multirow}
\usepackage{url}
\usepackage[linktocpage]{hyperref}


\hypersetup{
	colorlinks,
	linkcolor={red!60!black},
	citecolor={green!50!black},
	urlcolor={blue!80!black}
}


\def\nn{\nonumber}
\def \be{\begin{equation}}
\def \ee{\end{equation}}

\def \ba{\begin{align}}
\def \ea {\end{align}}
\def \l( {\left( }
\def \r) {\right)}
\def \RR {\mathcal{R}}
\def \co {compact object}

\def \df {dynamical friction}

\def \bomega {\tilde{\omega}}

\def \nn {\nonumber}

\def \f2om {f^2(\omega)}

\def \fR{\mathfrak{R}}

\begin{document}

\title{\bf Probing the quantum nature of black holes with ultra-light boson environments}


\author{Soumodeep Mitra}
\affiliation{Department of Physics, University of South Dakota, Vermillion, SD 57069, USA}

\author{Sumanta Chakraborty}
\affiliation{School of Physical Sciences, Indian Association for the Cultivation of Science, Kolkata 700032, India}

\author{Rodrigo Vicente}
\affiliation{Institut de Fisica d’Altes Energies (IFAE), The Barcelona Institute of Science and Technology, Campus UAB, 08193 Bellaterra (Barcelona), Spain}

\author{Justin C. Feng}
\affiliation{Leung Center for Cosmology and Particle Astrophysics, National Taiwan University, Taipei 10617, Taiwan}

\begin{abstract}
Quantum black holes (BHs), thought of as an excited multilevel system, can be effectively modelled by replacing an event horizon with a partially reflective membrane. This emergent feature affects their interaction with hosting environments, with the most pronounced effects happening for particles with mass~$m_{\rm p}\lesssim (10 M_{\odot}/M)\,10^{-11}\mathrm{\,eV}$, where~$M$ is the BH mass. We show that ultra-light bosons---a viable dark matter candidate---can be used to probe the quantum nature of BHs. We derive analytical expressions for the accretion rate and dynamical friction acting on exotic compact objects moving through an ultra-light scalar field, finding that while the accretion rate is sensitive to the quantum BH's reflectivity, the dynamical friction is the same as for classical BHs. We then use these expressions to estimate the orbital dephasing in the inspiralling of different binaries in the Laser Interferometer Space Antenna (LISA) band. Our results indicate that LISA may be able to discriminate quantum from classical BHs through their different accretion rates.
\end{abstract}

\maketitle

\section{Introduction}%

Black holes (BHs) behave in several respects as ordinary bound systems, which led many to argue that, in a theory of quantum gravity, their mass might be quantized. The idea of quantum BHs as giant (gravitational) \emph{atoms} was pioneered by~\citet{Bekenstein:1974jk} and~\citet{Mukhanov:1986me}, and led them via different paths  
to the same mass-spectrum: the horizon area must be quantized in multiples of the Planck area (see also~\cite{Hod:1998vk, Maggiore:2007nq, Dvali:2010gv, Dvali:2011nh}). If true, this would lead to imprints on the spectrum of Hawking radiation~\cite{Bekenstein:1995ju, Bekenstein:1997bt}, and also on the gravitational-wave (GW) signals of binary BH coalescences~\cite{Maselli:2017cmm, Maselli:2017cmm, Datta:2019epe, Cardoso:2019apo, Agullo:2020hxe, Sago:2021iku, Maggio:2021uge, datta2021imprint, Chakraborty:2022zlq, Chakraborty:2023zed, nair2023dynamical}. These are particularly distinctive for quanta of frequency~$f\lesssim (10 M_\odot/M)\mathrm{\,kHz}$ (with~$M$ the BH mass), which are within the range of current and near-future GW detectors. 
So, one is led to an astonishing conclusion: GW observations may be used to probe Planckian effects at the horizon scale~\cite{Maselli:2017cmm, Datta:2019epe, datta2021imprint, Chakraborty:2022zlq, Nair:2022xfm, Chakraborty:2023zed, nair2023dynamical, Datta:2024vll, Cardoso:2019apo, Agullo:2020hxe, Maselli:2017tfq, Abedi:2016hgu, Westerweck:2017hus, Nielsen:2018lkf, Lo:2018sep, Uchikata:2019frs}.

Although the current observations by the LIGO-Virgo-KAGRA collaboration are consistent with \emph{classical} BHs evolving in \emph{vacuum} GR~\cite{LIGOScientific:2019fpa, LIGOScientific:2020tif, LIGOScientific:2021sio, Fedrow:2017dpk, CanevaSantoro:2023aol}, they still leave open the possibility for exotic compact objects (ECOs)~\cite{Cardoso:2016rao, Mark:2017dnq, Correia:2018apm, Cardoso:2019rvt, Abedi:2022bph, Biswas:2023ofz} and the presence of astrophysical environments~\cite{Fedrow:2017dpk, CanevaSantoro:2023aol}. Extreme mass-ratio inspirals (EMRIs) and supermassive binary BHs---which will be seen by the space-borne LISA mission~\cite{Amaro-Seoane2017}---will be key in further constraining deviations to the classical BH paradigm (e.g., via tidal heating~\cite{Maselli:2017cmm, Datta:2019epe}) and probing environments (e.g., via dynamical friction (DF) and accretion~\cite{barausse2008influence, Caputo:2020irr, Kocsis:2011dr, barausse2014can, Cardoso:2019rou, kavanagh2020detecting, Coogan:2021uqv, Cole:2022yzw, Garg:2022nko, Speeney:2022ryg, Tomaselli:2023ysb, Brito:2023pyl, Duque:2023cac, Bertone:2024rxe}). But the two are not independent: most ECOs can be effectively modelled by replacing an event horizon with a partially reflective membrane~\cite{Cardoso:2019rvt}, affecting their interaction with environments and the dynamics within them. How effective are environments as \emph{discriminators} of compact objects?

For quantum BHs, the dynamics will be most strongly modified for environments of particles with mass~$m_\mathrm{p}\lesssim (10 M_{\odot}/M)\,10^{-11}\mathrm{\,eV}$, which is more than 16 orders of magnitude smaller than the electron mass. Thus, standard astrophysical environments (like accretion disks) are not suited to probe the quantum nature of BHs. But ultra-light boson particles can arise naturally in extensions to the Standard Model~\cite{arvanitaki2010string, Freitas:2021cfi} and are viable dark matter candidates (see, e.g.,~\cite{hui2017ultralight, ferreira2021ultra, Hui:2021tkt, fabbrichesi2021physics}). Most remarkably, dense environments of bosons with such masses can form efficiently around BHs (e.g., via superradiance~\cite{Brito:2015oca, Baumann:2019eav, Hui:2022sri}, accretion~\cite{Hui:2019aqm, Clough:2019jpm, Bamber:2020bpu}, or dynamical capture~\cite{Budker:2023sex}). In this paper, we will show that LISA may be able to discriminate quantum from classical BHs through the environmental effects of ultra-light bosons.

The paper is organized as follows: in Sec.~\ref{refQBH} we discuss the basics of the models of quantum BHs that we considered. Ultra-light boson environments and their effects are studied subsequently in Secs.~\ref{ultralightboson} and~\ref{environment}, respectively. In Sec.\ref{dynQBH} we discuss the dynamics of these quantum BHs in ultra-light boson environments, while the implications for future GW observations are finally presented in Sec.~\ref{ObsPro}. We conclude with a discussion of our results. Some more detailed calculations are presented in the appendices.

\noindent\emph{Notations and conventions:} we use $G=c=k_{\rm B}=1$ and the mostly positive metric signature.

\section{Reflectivity of quantum BHs}\label{refQBH}
In this work, quantum BHs will be thought of as excited multi-level systems that are slowly decaying through Hawking radiation at temperature
\begin{equation}
T_{\rm H}\equiv \frac{\hbar\kappa}{2\pi}~,
\qquad
\kappa=\frac{\sqrt{1-\chi^2}}{2M(1+\sqrt{1-\chi^2})}~,
\end{equation}
for a Kerr BH with spin~$\chi\equiv (J/M^2)$. Simple heuristic arguments (e.g.,~\cite{Bekenstein:1974jk, Mukhanov:1986me, Hod:1998vk, Maggiore:2007nq, Dvali:2010gv, Dvali:2011nh}) suggest that axisymmetric quantum BHs may have their area quantized as~$A_{\rm h}\approx N\upalpha \ell_{\rm P}^2$ (for~$N\gg1$) and spin as~$J_z=j \hbar$, where $\upalpha\sim O(1)$ is a parameter depending on the quantum gravity model,~$\ell_{\rm P}=\sqrt{\hbar}$ is the Planck length, and~$N$ and~$j\leq (\upalpha N/8\pi)$ are positive integers; this idea has also been realised in several quantum gravity schemes (see, e.g.,~\cite{Kogan:1994fg, Maggiore:1994ww, Lousto:1994jd, Peleg:1995gg, Louko:1996md, Barvinsky:1996hr, Perez:2017cmj}). This results in the BH mass-spectrum
\begin{equation}
\frac{M_{N,j}}{\ell_{\rm P}}\approx\sqrt{\frac{\upalpha N}{16 \pi}+\frac{4\pi j^2}{N \upalpha}}\,,
\end{equation}
implying that BHs can only absorb quanta~$(\omega_{}, l, m)$ satisfying
\begin{equation}
\hbar\omega_{n,m}  = \frac{\upalpha}{4} T_{\rm H} \,n + \hbar\Omega_ {\rm h}\, m +O(N^{-1})~,
\end{equation}
with~$n=\Delta N$ and~$m=\Delta j$, where~$\Omega_ {\rm h}\equiv (\kappa\chi/\sqrt{1-\chi^2})$. Due to the \emph{spontaneous} decay generating Hawking radiation, the absorption lines have a width~$\Gamma(M, \chi)$, which can be computed numerically using the approximation scheme of Ref.~\cite{Agullo:2020hxe}, based on Page's computation~\cite{Page:1976ki} (see also~\cite{Hod:1998vk, Coates:2019bun}). Here, we shall use the fit expression for the line-width~$\Gamma(M, \chi)$ presented in Eq.~(5) of Ref.~\cite{datta2021imprint}.

Not knowing the details of the BH microphysics, one might consider the minimal model in which quantum BHs totally absorb the impinging modes associated with an allowed transition and totally reflect the others; so, one can take the reflectivity of the area-quantized BHs as~\cite{datta2021imprint}
\begin{equation}\label{AQB_reflectivity}
(\fR_{\rm A})_l^m=1-\sum_{n} W(\omega - \omega_{n,m}, \tfrac{\Gamma}{2}),
\end{equation}
with~$W(x,y>0)$ the Hann window function (which has support on~$|x|<y$). 

One should also consider the probability of absorption for each allowed transition and the probabilities of stimulated and spontaneous (Hawking) emission for the reverse transition. 
Indeed, using detailed balance (as in Ref.~\cite{oshita2020reflectivity}), one can show that these considerations are responsible for a suppression factor~$[1-g_N/g_{N+n}]$ in the BH's absorptance, with~$g_N$ being the degeneracy of the state~$(N,j)$. As noted by~\citet{Mukhanov:1986me}, the entropy change~$\Delta S=\Delta A_{\rm h}/4 \hbar$ implies
\begin{equation}
\frac{g_N}{g_{N+n}}\approx \exp\left[-\frac{\hbar |\bomega_{n,m}|}{T_{\rm H}}\right]\,,
\end{equation}
with the near-horizon frequency~${\bomega}_{n,m}\equiv \omega_{n,m}-m\Omega_{\rm h}$; thus, more careful modelling of area-quantized BHs may include an additional Boltzmann factor in Eq.~\eqref{AQB_reflectivity}. 
However, a Boltzmann-suppressed reflectivity can also arise in other models (e.g., from dissipation via the interaction with quantum/thermal fields close to the BH horizon~\cite{oshita2020reflectivity}); thus, for completeness, we will consider the effect of Boltzmann reflectivity separately, with
\begin{equation}
\label{QBH_reflectivity}
(\fR_{\rm B})_l^m=\exp\left[-\frac{\hbar\bomega}{T_{\rm H}}\right]~,
\end{equation}
where, $\bomega$ is the near-horizon frequency and $T_{\rm H}$ is the Hawking temperature, both of which have been defined earlier. 

As we will shortly discuss, for ultra-light bosons such that~$\omega M \ll1$, only the~$l=m=0$ modes can reach the strong-field region, resulting in a \emph{selection rule} for the allowed transitions. This implies an equidistant line spacing~$\hbar \Delta\omega=(\upalpha T_{\rm H}/4)$, which is larger than the line-width~$\Gamma$ for~$\chi\lesssim 0.8$ and~$\upalpha\gtrsim 1$. On the contrary, any Standard Model massive particle has~$\omega M \gg 1$ (for astrophysical BHs), resulting in~$l\gg 1$ modes reaching the strong-field region.\footnote{The critical impact parameter for particle capture is~$b_{\rm cr}\sim M$. In the eikonal ($l\gg1$) limit:~$b\approx (l/\omega v)$, with~$v$ the BH velocity. Thus,~$l_{\rm cr}\sim \omega M v$. E.g., for electrons:~$l_{\rm cr}\sim 10^{17} v (M/10 M_\odot)$.}
The absence of selection rules leads to overlap between highly excited states, blurring the discreteness of quantum BHs, as noted in Ref.~\cite{Agullo:2020hxe} (see also Refs.~\cite{Hod:2015qfc, Coates:2019bun}). The effect from the Boltzmann factor is also washed out for such particles (since~$e^{-\hbar\bomega/T_{\rm H}}\ll 1$). Thus, in this letter, we focus on ultra-light bosons, which are especially suited to probe the quantum nature of BHs.

\section{Ultra-light boson environment}\label{ultralightboson}
To model the interaction of ultra-light bosons with a quantum BH, we consider a massive complex scalar field~$\Phi$ satisfying the Klein-Gordon equation
\begin{equation}
(\Box_{\bm g} -\mu^2)\Phi=0~,
\end{equation}
for particles of mass~$m_{\rm p}\equiv \hbar \mu$, along with the covariant d'Alembertian~$\Box_g\equiv g^{\mu \nu}\nabla_\mu \nabla_\nu$, and whose energy-momentum tensor is
\begin{equation}\label{EMTDM}
T_{\mu \nu}=\nabla_{(\mu} \Phi^* \nabla_{\nu)} \Phi-\frac{1}{2}g_{\mu \nu}\left(\nabla^\alpha \Phi^*\nabla_\alpha \Phi+\mu^2 |\Phi|^2\right)~.
\end{equation}
We assume that the environment is dilute enough that its backreaction on the spacetime geometry is negligible, and consider that the exterior of a quantum BH is described by a stationary and axisymmetric Kerr metric. We focus on a homogeneous environment with rest-mass density~$\rho_0$, and neglect any velocity dispersion of the particles in the medium; in the BH rest frame, the asymptotic field is
\begin{equation}
\Phi(r\to \infty)\approx \sqrt{\frac{\rho_0}{\mu}}\,e^{-i (\omega t-k_{\infty}^i x_i)}~,
\end{equation}
with~$k_\infty^i=k_\infty \xi^i$, where~$k_{\infty}\equiv\sqrt{\omega^2-\mu^2}$ and~$\xi^i$ is a unit-vector. Note that~$v^i\equiv-k^i_\infty/\omega$ is the velocity of the BH relative to the medium; or, equivalently,~$\omega=\gamma \mu$ and~$k_\infty^i=-\gamma \mu v^i$ with the Lorentz factor~$\gamma \equiv 1/\sqrt{1-v^2}$. 

To find the (stationary) field distribution around the quantum BH we start by first decomposing the field as
\begin{equation}\label{scalardecomp}
\Phi = \sum_{l,m} e^{-i \left( \omega t - m \varphi \right)} \mathrm{\,Ps}_{l}^{m}(\cos \theta,\upgamma^2) \RR_l^m(r)\,,
\end{equation}
with~$\mathrm{\,Ps}_{l}^{m}$ being the (oblate) angular spheroidal wave functions of the first kind~\cite{NIST:DLMF} with~$\upgamma\equiv i a k_{\infty}$, and radial profiles satisfying
\begin{align}
\label{master_wave}
&\Delta \frac{d}{dr} \left [\Delta \frac{d \RR_{l}^{m}}{dr} \right]+\Big[\omega^2 (r^2 + a^2)^2 -4am\omega M r 
\nonumber
\\
&\quad +(ma)^2 -\left(\lambda^m_l + \mu^2 (r^2 +a^2)\right) \Delta  \Big] \RR_{l}^{m}=0~,
\end{align}
with~$a\equiv (J/M)$,~$\Delta \equiv r^2 + a^2-2Mr$, and~$\lambda_l^m$ (with~$l\geq|m|$) the eigenvalues from the angular equation, which do not possess a closed-form expression.
The asymptotic boundary condition implies 
\begin{equation}\label{BC_infinity}
\mathcal{R}_{l}^{m}(r\to \infty)\approx r^{-1}(I_{l}^{m}e^{-i \phi_\infty(r)}+R_{l}^{m} e^{i \phi_\infty(r)})\,, 
\end{equation}
with $\phi_\infty \equiv k_\infty r- \eta \ln(2 k_\infty r)$ and~\cite{vicente2022dynamical}
\begin{equation} \label{incident_ampl}
I_{l}^{m}=\frac{2l+1}{2k_{\infty}} \frac{\sqrt{\rho_0}}{\mu}\frac{(l-m)!}{(l+m)!}(-i)^{m+1} {\rm \, Ps}_l^m(\cos\vartheta,\gamma^2)~,
\end{equation}
where~$\eta \equiv-\{\alpha_{\rm g}(1+v^2)/v\}$ and~$\xi^i\equiv(0,\sin \vartheta,-\cos \vartheta)$,\footnote{We work in asymptotic Cartesian-like coordinates, such that the spin of the quantum BH is aligned along $(0,0,1)$.} with the gravitational coupling~$\alpha_{\rm g}\equiv \gamma \mu M$. The difference to a classical BH originates from the field distribution at~$r_\epsilon =r_{\rm h}(1+\epsilon)$, with~$0<\epsilon\ll1$ and the location of the BH horizon being~$r_{\rm h}\equiv M(1+\sqrt{1-\chi^2})$, which includes now a reflected part,
\begin{equation}\label{BC_horizon}
\RR_{l}^{m}(r\to r_\epsilon) \approx \frac{T_{l}^{m}}{\sqrt{r_\epsilon^{2}+a^{2}}}
\Big[e^{-i\bomega r_{*}} + \sqrt{\fR_{l}^{m}} e^{i(\bomega r_{*}+\psi_l^m)}\big]\,,
\end{equation}
where~$(dr/dr_*)=\Delta/(r^2+a^2)$ and the phase-shift~$\psi_l^m$ is a model-dependent quantity~\cite{Zimmerman:2023hua}. For ultra-light bosons, such that~$\alpha_{\rm g} \ll1$, Eq.~\eqref{master_wave} can be solved approximately using asymptotic matched expansions~\cite{Starobinsky:1973aij, unruh1976absorption}. Note that this method applies for computing the scalar field distribution around any ECOs, and not just quantum BHs. 
%

\section{Environmental effects}\label{environment}
To compute the accretion rate and gravitational drag acting on quantum BHs, we extended the approach of Ref.~\cite{vicente2022dynamical} to ECOs. When the steady regime is attained, these effects can be characterised entirely by the scalar field asymptotics~\cite{vicente2022dynamical, Clough:2021qlv}. In the ECO frame, one has (see Appendix \ref{appnA} for derivation)
\begin{equation}
\dot{M}\approx \lim_{r \to \infty} r^2 \int T_{t r} d\Omega_2
\end{equation}
and
\begin{equation}
F_i\approx \lim_{r \to \infty} r^2\int T_{r i} d\Omega_2,
\end{equation}
where~$d\Omega_2$ is the volume element on the 2-sphere. Note that $T_{tr}$ is related to the radial flux of matter energy-momentum tensor and hence is responsible for the accretion of dark matter mass, leading to non-zero $\dot{M}$. Similarly, the $T_{ri}$ components of the matter energy-momentum tensor are related to the force exerted by the dark matter particles along the $i$th direction and contributes to the force arising out of dynamical friction.   

The above expressions for the rate of accretion and the frictional force can be determined in terms of the radial function $\mathcal{R}_{l}^{m}$, given the expression for the energy momentum tensor in Eq. \eqref{EMTDM} and the decomposition of the scalar field in Eq. \eqref{scalardecomp}. Given the asymptotic behaviour of the radial function in Eq. \eqref{BC_infinity}, and for~$\alpha_{\rm g} \ll1$, the above exercise yields the following expressions for the accretion rate and the dynamical friction force in the frame of the BH (see Appendix \ref{appnB} and Appendix \ref{appnC} for a detailed derivation):
\begin{subequations}
    \begin{gather}
        \dot{M}\approx \rho M^2  \left(\frac{\pi v}{\alpha_{\rm g}^2}\right)\sum_l (2l+1) \left(1-\left|\frac{R_l^0}{I_l^0}\right|^{2}\right)~,
        \\
        F^i \approx -  \rho M^2 \xi^i \left(\frac{2 \pi}{\alpha_{\rm g}^2}\right) \sum_l (l+1){\rm Re}\left[1+\left(\frac{R_l^0}{I_l^0} \right)^* \frac{R_{l+1}^0}{I_{l+1}^0}\right]~,
    \end{gather}
\end{subequations}
where~$\rho\equiv \rho_{0}\gamma^2$ is the environment density in the ECO frame.
However, we want to express these quantities in terms of the reflectivity of the horizon for quantum BHs, and ECOs, in general. 
For this purpose we need to solve for the radial function in different limits and match the corresponding expressions. 
This is achieved by the following route --- (a) solving Eq. \eqref{master_wave} near the horizon and then imposing Eq. \eqref{BC_horizon} as the boundary condition, (b) solving Eq. \eqref{master_wave} in the asymptotic regime and then imposing Eq. \eqref{BC_infinity} as the boundary condition, and finally (c) matching the asymptotic limit of the near-horizon solution with the near-horizon limit of the asymptotic solution. 
Thereby relating the ratio of the reflected and the incident wave at infinity ($R_{l}^{m}/I_{l}^{m}$) with the reflectivity $\fR_{l}^{m}$ at the horizon. 
The above matched asymptotic expansions and the relation between the ratio of scattering amplitudes~$(R_l^m/I_l^m)$ with the reflectivity has been presented in Appendix \ref{appnB} and Appendix \ref{appnC}, respectively. We quote the final results here:
\begin{subequations}
\begin{gather}
    \dot{M} \approx \rho A_{\rm h} \left(\frac{e^{-\pi \eta}\pi \eta}{\sinh (\pi \eta)} \right)\mathcal{T}(\alpha_{\rm g},\chi)~,
    \label{accret}
    \\
    F^i \approx -\frac{4 \pi \rho M^2}{v^2}(1+v^2)^2 \xi^i\mathcal{D}(\eta,\Lambda) -\dot{M} v^i, 
    \label{drag_force}
\end{gather}
\end{subequations}
where
\begin{equation}
\mathcal{T}\equiv{\rm Re}\left[\frac{(1-e^{i \psi_0^0}\sqrt{\fR_0^0})}{(1+e^{i \psi_0^0}\sqrt{\fR_0^0})}\right]~,
\end{equation}
and~$\mathcal{D}\equiv\{ \ln \Lambda- {\rm Re} [\Psi(1- i \eta)]\}$, with~$\Psi$ the digamma function, and~$\Lambda\equiv \gamma v \mu b_{\rm max}\gg1$ the ratio of the maximum impact parameter,~$b_{\rm max}$,\footnote{The maximum impact parameter~$b_{\rm max}\equiv \sqrt{l_{\rm max}(l_{\rm max}+1)}/k_\infty$ must be introduced to regularise a logarithmic (infrared) divergence originating from the long range~$\propto 1/r$ character of gravity.} to the reduced de Broglie wavelength, $\lambdabar_{\rm dB}\equiv (\gamma \mu v)^{-1}$. When~$\fR_0^0=0$, these expressions reduce to the ones derived for classical BHs in Ref.~\cite{vicente2022dynamical} shown to describe well numerical simulations in Refs.~\cite{Traykova:2021dua, Traykova:2023qyv}. For~$v\gtrsim 2 \mu M$, it can be shown that~$\mathcal{D}\approx {\rm Cin}\, \Lambda_j+(\sin \Lambda_j/\Lambda_j)-1$, for any~$\Lambda$~\cite{hui2017ultralight, Traykova:2023qyv}, with~${\rm Cin} \,x \equiv \int_0^x d\bar{x} (1-\cos \bar{x})/\bar{x}$.

Due to their ultra-light mass, the scalars cannot probe the strong-field region and the DF (first term in the drag force) is not sensitive to the nature of the compact object; this explains the independence on~$\fR_l^m$ and the similarity to the Newtonian expression derived in~\cite{hui2017ultralight} (modulo kinematic relativistic corrections). Instead, the accretion rate and associated momentum transfer (second term in the drag force) must sense the strong-field region. Nevertheless, due to the ultra-light mass ($\alpha_{\rm g} \ll1$), only the mode~$l= 0$ penetrates significantly the potential barrier; thus, at leading order in~$\alpha_{\rm g}$, the environmental effects are sensitive only to~$\fR_0^0$.

\section{Dynamics of quantum BHs}\label{dynQBH}
An asymptotic observer at rest relatively to the medium perceives the BH's rest mass changing at a rate~$\dot{M}'=\dot{M}/\gamma$ and its linear momentum at~$F_i'=F_i+\dot{M} v^i$. Thus, from Eq.~\eqref{accret} and Eq.~\eqref{drag_force}, it follows that at leading order in~$\alpha_{\rm g}$, only the accretion rate contains information about the nature of the ECO. Accretion affects the dynamics of binaries and their corresponding GW waveform by imparting a dephasing on their signal~\cite{barausse2014can}; here, we compare said effect for classical and quantum BHs. 
 
We focus on the dominant~$m=2$ mode of the GW waveform associated with quasi-circular inspirals. The interval of time corresponding to a chirp in the signal's frequency from~$v_{\rm i}$ to~$v_{\rm f}$, where $v\equiv(\pi M_{\rm T} f)^{1/3}$, is
\begin{equation}
\Delta t=-\int_{v_{\rm i}}^{v_{\rm f}}d \bar{v}\, \frac{\mathcal{E}'(\bar{v})}{\mathcal{F}(\bar{v})}~,
\end{equation}
over which the signal's phase changes by
\begin{equation}
\upvarphi=-\frac{2}{M_{\rm T}}\int_{v_{\rm i}}^{v_{\rm f}}d \bar{v}\,\bar{v}^3 \frac{\mathcal{E}'(\bar{v})}{\mathcal{F}(\bar{v})}~, 
\end{equation}
with $M_{\rm T}$ the total binary mass, $\mathcal{E}$ the orbital energy, and~$\mathcal{F}$ the dissipation rate (via GW emission, accretion, and other dissipation mechanisms, like drag forces, tidal heating~\cite{Datta:2019epe, datta2021imprint, Chakraborty:2021gdf}, tidal deformability~\cite{Maselli:2017cmm, Chakraborty:2023zed}).
Though one could easily include higher-order post-Newtonian (PN) corrections, in this estimate we consider just the leading-order contributions. Then, assuming that the binary components move at non-relativistic velocities, it is straightforward to find
\begin{equation} \label{integr_phase}
    \frac{\mathcal{E}'}{\mathcal{F}} \approx -\frac{\upeta M_{\rm T} v}{\mathcal{F}_{\rm GW}}\Big[1- \mathcal{F}_{\rm GW}^{-1}\big( v^2[2\upeta\dot{M}_{\rm T}+\dot{\upeta} M_{\rm T}]+\mathcal{F}_{\rm diss}\big)\Big]\,,
\end{equation}
with the symmetric mass-ratio~$\upeta\equiv M_1 M_2/M_{\rm T}^2$~($\leq 1/4$).\footnote{We used that the orbital angular momentum is conserved in the accretion process (see App. of~\cite{Caputo:2020irr}), and assumed that the scalar environment is rotating much more slowly than the binary.}

We are interested in assessing the effectiveness of accretion in distinguishing quantum from classical BHs. So, we will neglect the dephasing due to tidal heating/deformability---which, as shown in Refs.~\cite{Maselli:2017cmm, Datta:2019epe, Sago:2021iku, Maggio:2021uge, datta2021imprint, Chakraborty:2023zed}, may also be used for the same purpose. Even though DF is not sensitive to the strong-field region, we include its effect on the GW phase (as it is generally the dominant environmental effect~\cite{barausse2014can, Cardoso:2019rou}). So, from the first term in Eq.~\eqref{drag_force}, we take 
\begin{equation*}
\mathcal{F}_{\rm diss}\approx 4\pi \rho_0\left[\frac{M_1^2 \mathcal{D}_1}{v_1}+\frac{M_2^2 \mathcal{D}_2}{v_2}\right]\,, 
\end{equation*}
with~$b_{{\rm max},j}\approx 2M_{\rm T}(v_j/v^3)$,\footnote{A (heuristic) extrapolation from linear to circular motion can be done by replacing~$b_{{\rm max},i}\sim 2 r_i$ (see, e.g., App. D of~\cite{hui2017ultralight} or~\cite{vicente2022dynamical}). Analytical (Newtonian) expressions for the DF in circular motion were derived in Ref.~\cite{Buehler:2022tmr}.} where~$r$ is the orbital separation distance, and~$\Lambda_j\approx 2 \mu M_{\rm T}(v_j^2/v^3)$. 

For a field configuration sustained by a BH binary system (or by the primary of an EMRI), one expects~$\mu M_{\rm T}\lesssim 0.1$ [i.e., $m_{\rm p}\lesssim 10^{-17}\,{\rm eV}\,(10^6 M_\odot/M_{\rm T})$]. If the configuration is sourced by superradiance~\cite{Zhou:2023sps}, it may saturate in a scalar field with total mass~$\sim 0.1 M_{\rm T}$~\cite{East:2017ovw, Herdeiro:2021znw}, distributed over a region of radius $\sim 1/(\mu^2 M_{\rm T})$; thus, with an average density $\rho_0 M_{\rm T}^2 \sim 0.1(\mu M_{\rm T})^6$. Note that, in the LISA band, typical velocity of the secondary corresponds to, 
\begin{equation}
v\sim 0.249\left(\frac{M_{\rm T}}{10^6 M_\odot}\right)^{1/3}\left(\frac{f}{1 {\rm\, mHz}}\right)^{1/3}~,
\end{equation}
and~$v<v_{\rm isco}\sim 1/\sqrt{6}\approx 0.4$. We focus on~$v>\mu M_{\rm T}$, so that the binary separation is smaller than~$1/(\mu^2 M_{\rm T})$.

Let us first consider the case of an EMRI in the LISA band, with mass-ratio~$q\equiv M_1/M_2 \gg 1$. We use~$v_1\approx 0$ and~$v_2 \approx v$. While it is easy to check that~$(\mathcal{D}_1/v_1)\approx 0$, the primary's accretion rate~$\dot{M}_1\propto \mu M_1/v_1$ seems to diverge as~$v_1\to 0$. However, note that this expression describes a steady state, whose characteristic length scale is~$M_1/v_1^2$. In the massive BH rest frame, the perturbations in the environment propagate with group velocity~$\partial \omega_1/\partial k_{\infty,1}=k_{1,\infty}/\omega_1\sim v_1$, and so the steady-state is attained over a timescale~$\sim M_1/v_1^3$. We can anticipate that a time-dependent treatment (as, e.g., in Refs.~\cite{ostriker1999dynamical, Vicente:2019ilr}) would result in the replacement of~$\mu M_1/v_1$ by a term~$\sim\mu v_1^2 t$, which vanishes as~$v_1\to 0$. So, we take~$\dot{M}_1\approx0$.

Therefore, will all of these inputs, for an EMRI in the LISA band, the ratio $(\mathcal{E}'/\mathcal{F})$, as presented in Eq.~\eqref{integr_phase} of the main text reduces to,
\begin{equation}
\frac{\mathcal{E}'}{\mathcal{F}} \approx -\frac{5 q M_{\rm T}}{32 v^9}\Big\{1- \frac{5\pi \rho_0 M_{\rm T}^2}{8 v^{11}}\big[\mathcal{D}_2+2v^3\tfrac{r_{{\rm h},2}}{M_2} \mathcal{T}_2\big]\Big\}~,
\end{equation}
with
\begin{align}
\mathcal{T}_2=\frac{1- \fR_2}{1+2 \sqrt{\fR_2} \cos \psi_2+\fR_2}\,,
\\
\nn
\end{align}
being a function of~$\mu M_2$ and $\chi_2$\footnote{To avoid clutter we omit the~$l=m=0$ indices in $\fR$ and $\psi$.}. Since~$\Lambda_2\lesssim 1$, we use~$\mathcal{D}_2\approx (\Lambda_2^2/12) \approx (\mu M_{\rm T}/\sqrt{3}v)^2$. For a fiducial EMRI with~$M_{\rm T}\gtrsim 10^6 M_\odot$ and~$f\gtrsim 1{\rm \, mHz}$, the condition~$v >\mu M_{\rm T}$ implies a maximal coupling~$\mu M_{\rm T} \lesssim 0.249$ and corresponding average density $\rho_0 M_{\rm T}^2\lesssim 10^{-5}$ [i.e., $\rho_0\lesssim 6\, \mathrm{g/cm^{3}}(10^6 M_\odot/M_{\rm T})^2$].
For an observation period $T_{\rm obs}  \sim 4\,{\rm yr}$, the chirp signal's frequency changes as: $v_{\rm i}^{-8}-v_{\rm f}^{-8}\approx 2^8\, T_{\rm obs}/(5q M_{\rm T})$;\footnote{We neglect the environmental effects in this relation. This is a conservative estimate, since these effects tend to accelerate the inspiral (i.e., lead to larger~$\Delta v$ for same~$T_{\rm obs}$).} so, the change of velocity becomes,
\begin{equation}
\Delta v \approx 0.006\left(\frac{v}{0.25}\right)^9 \left(\frac{T_{\rm obs}}{4\, {\rm yr}}\right)\left(\frac{10^5}{q}\right)\left(\frac{10^6M_\odot}{M_{\rm T}}\right)~,
\end{equation}
for quasi-monochromatic signals (i.e., $\Delta v\ll v$). Thus the phasing can be decomposed as~$\upvarphi \approx \upvarphi^{\rm vac}+\upvarphi^{\rm df}+\upvarphi^{\rm accr}$, where 
\begin{align}
\upvarphi^{\rm vac}\sim 8 \times10^5 \left(\frac{f}{1\, {\rm mHz}}\right) \left(\frac{T_{\rm obs}}{4 \, {\rm yr}}\right)\,,
\end{align}
is the phase shift coming from the Newtonian description in the absence of any environmental effect. On the other hand, the contribution from the dynamical friction term to the phasing of the GW from an EMRI reads,
\begin{align}
\upvarphi^{\rm df}&\sim-3\times 10^{3}\left(\frac{\rho_0 M_{\rm T}^2}{10^{-8}}\right)\left(\frac{\mu M_{\rm T}}{0.1}\right)^2 
\nonumber
\\
&\qquad \times\left(\frac{10^6 M_\odot}{M_{\rm T}}\right)^\frac{13}{3}\left(\frac{1\,{\rm mHz}}{f} \right)^{\frac{10}{3}} \left(\frac{T_{\rm obs}}{4\,{\rm yr}}\right)\,,
\end{align}
and for accretion the phasing formula takes the following form,
\begin{align}
\upvarphi^{\rm acc}&\sim -2 \times10^3\,\mathcal{T}_2 \left(\frac{\rho_0 M_{\rm T}^2}{10^{-8}}\right) \left(\frac{10^6 M_\odot}{M_{\rm T}} \right)^\frac{8}{3}
\nonumber
\\
&\qquad \times\left(\frac{1\,{\rm mHz}}{f} \right)^{\frac{5}{3}} \left(\frac{T_{\rm obs}}{4 \, {\rm yr}}\right)\,.
\end{align}
This concludes our discussion involving the effect of dynamical friction and accretion arising from ultra-light boson environment for EMRI. 

Let us turn our attention now to an equal-mass binary with $q\approx 1$. By symmetry, one finds that $\dot{\eta}=0$ and $v_{1}\approx v_{2}\approx (v/2)$. For such a binary Eq.~\eqref{integr_phase} in the main text becomes\footnote{We consider binary components with equal spin~$\chi_1=\chi_2=\chi$, so that it follows: $\mathcal{T}_{1}=\mathcal{T}_{2}=\mathcal{T}$. }
\begin{equation}
\frac{\mathcal{E}'}{\mathcal{F}} \approx -\frac{5 M_{\rm T}}{8 v^9}\Bigg\{1-\frac{10\pi \rho_0 M_{\rm T}^2}{v^{11}}\Big[\mathcal{D}_1+v^3\frac{r_{{\rm h},1}}{M_{\rm T}} \mathcal{T}_1\Big]\Bigg\}~,
\end{equation}
with~$\mathcal{D}_1 \approx (\Lambda_1^2/12) \approx (\mu M_{\rm T}/4\sqrt{3}v)^2=\mathcal{D}_2$, since~$\Lambda_1=\Lambda_{2}\lesssim 1$. For a massive system ($M_{\rm T}\gtrsim 10^5 M_\odot$) in the LISA band, the signal is far from monochromatic. So, we take $v_{\rm f}=(v_{\rm isco}/2)\approx 0.2$ and 
\begin{equation}
v_{\rm i}=\max\Big\{\mu M_{\rm T},\big[v_{\rm f}^{-8}+2^8\, (T_{\rm obs}/5 M_{\rm T})\big]^{-\frac{1}{8}}\Big\}~,
\end{equation}
where the second element is~$\gtrsim 0.07$ for $M_ {\rm T}\gtrsim 10^6 M_{\odot}$ and~$T_ {\rm obs}=4\, {\rm yr}$. From~$v_{\rm i}$ to~$v_{\rm f}$, the phasing is 
\begin{align}
    &\upvarphi^{\rm vac} \approx \frac{v_{\rm f}^5-v_{\rm i}^5}{4 (v_{\rm i} v_{\rm f})^5}~, 
    \\ 
    &\upvarphi^{\rm df} \approx -\frac{25 \pi (v_{\rm f}^{18}-v_{\rm i}^{18})}{1728 (v_{\rm i} v_{\rm f})^{18}} \rho_0 M_{\rm T}^2(\mu M_{\rm T})^2~,
    \\
    &\upvarphi^{\rm acc}\approx  -\frac{25 \pi \mathcal{T} (v_{\rm f}^{13}-v_{\rm i}^{13})}{52 (v_{\rm i}v_{\rm f})^{13}}\rho_0 M_{\rm T}^2(1+\sqrt{\smash[b]{1-\chi^2}})~.
\end{align}
Note that our treatment holds for densities small enough that the environmental effects are but a small perturbation to the GW radiation-driven inspiral. Interestingly, as shown in Fig.~\ref{fig:maxdensity}, for couplings~$\mu M_{\rm T}\gtrsim 0.03$, these are much more diluted than the maximal~$\rho M_{\rm T}^2 \sim 0.1 (\mu M_{\rm T})^6$. The minima in the coloured curves of Fig.~\ref{fig:maxdensity} signal the transition to~$T_{\rm obs}< 4\,{\rm yr}$. For larger couplings~$\alpha_{\rm g}$, the scalar field configuration becomes small enough that it takes less than~$4\,{\rm yr}$ from the moment the binary enters the overdensity, at~$v_{\rm i}\approx \mu M_{\rm T}$, to the moment it reaches~$v_{\rm f}=v_{\rm isco}/2$.

\section{Observational prospects}\label{ObsPro}
Many works indicate that future GW observations will be able to detect environmental effects from dark matter overdensities on the GW signal of binary inspirals (e.g,~\cite{Baumann:2018vus, Baumann:2019eav, Cardoso:2019rou, kavanagh2020detecting, Coogan:2021uqv, Cole:2022yzw, Tomaselli:2023ysb, Brito:2023pyl, Duque:2023cac, CanevaSantoro:2023aol, Tomaselli:2024bdd, Boskovic:2024fga, Bertone:2024rxe}). Can such observations also be used to tell the difference between classical and quantum BHs (or, more generally, any ECOs)? 

Using a Fisher (information) matrix analysis, it can be shown that with LISA we will be able to estimate the GW frequency of quasi-monochromatic sources with a relative error~\cite{Takahashi:2002ky}
\begin{equation*}
\frac{\delta f}{f} \approx 5 \times 10^{-3}\, \left(\frac{10^2}{{\rm SNR}}\right)\left(\frac{4\,{\rm yr}}{T_{\rm obs}}\right)~.
\end{equation*}
Following which, here we assume that LISA will be sensitive to phase changes, such that, 
\begin{equation}
\frac{\delta \upvarphi}{\upvarphi^{\rm vac}} \sim 5 \times 10^{-3} \, \left(\frac{10^2}{{\rm SNR}}\right)\left(\frac{4\,{\rm yr}}{T_{\rm obs}}\right)~.
\label{boundphi}
\end{equation}
Even though this assumption is less justified for ``chirping'' signals, we still take it for an order of magnitude estimate (the same assumption was taken, e.g., in Ref.~\cite{barausse2014can}).

\begin{figure}[t]
    \centering
    \includegraphics[width = 0.99\linewidth]{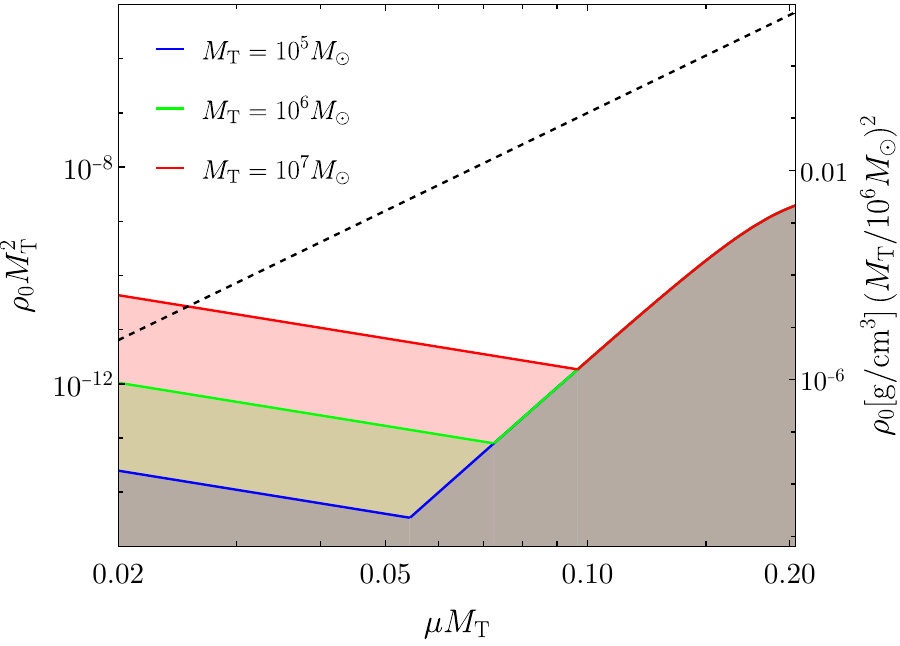}
    \caption{Solid lines: density for which~$|\upvarphi-\upvarphi^{\rm vac}|/ \upvarphi^{\rm vac}=0.05$, which we take as the boundary of the regime of validity of our perturbative treatment of environmental effects. Dashed: maximal density for clouds with total mass~$\sim 0.1 M_{\rm T}$. LISA may be sensitive to densities~$\gtrsim 10^{-3} \, (10^4/{\rm SNR})(4\,{\rm yr}/T_{\rm obs})$ times smaller than the coloured solid lines.}
    \label{fig:maxdensity}
\end{figure}

This estimate indicates that LISA will be capable of detecting the effect of DF from ultra-light bosons for (with $T_{\rm obs}=4\textrm{yr}$)
\begin{align}
\rho_0 &\overset{(q\gg1)}{\gtrsim} 10^{-3}\, \mathrm{g/cm^3} \left(\frac{0.1}{\mu M_{\rm T}}\right)^2\left(\frac{M_{\rm T}}{10^6 M_\odot} \right)^{\frac{7}{3}}
\nonumber
\\
&\qquad \times\left(\frac{f}{1\,{\rm mHz}}\right)^{\frac{13}{3}}\left(\frac{10^3}{{\rm SNR}}\right)~,
\end{align}
with EMRIs, while symmetric massive binaries merging in the LISA band could, in principle, be used to probe boson densities larger than~$10^{-3} \, (10^4/{\rm SNR})(4\,{\rm yr}/T_{\rm obs})$ times the ones shown in the solid coloured curves of Fig.~\ref{fig:maxdensity}. Our estimates for supermassive BH binaries are in remarkable agreement with the findings of Refs.~\cite{Cardoso:2019rou, Brax:2024yqh} obtained with a Fisher matrix analysis, but disagree by an astonishing~$13$ orders of magnitude for EMRIs; the latter can be traced back to their (invalid) assumption of~$\mathcal{D}_1\approx 1$, leading to an extra factor~$q^3$ in $\upvarphi^{\rm df}$.\footnote{The $q=10^4$ case considered there, would lead to $\sim q^3=10^{12}$ smaller detectable densities.}

\begin{figure}[t]
    \centering
    \includegraphics[width = 0.95\linewidth]{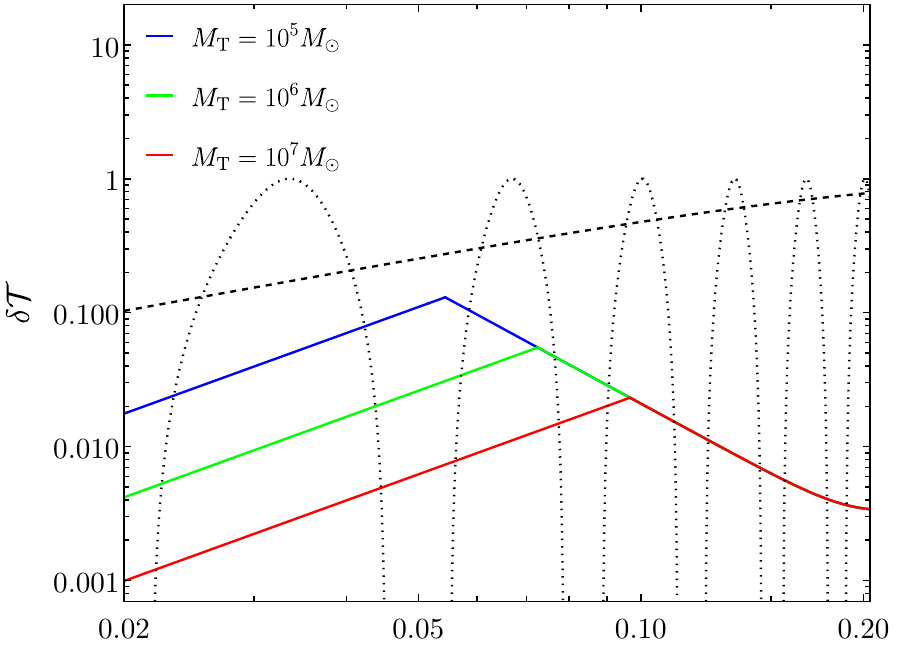}
    \caption{Solid lines: Estimate of~$\mathcal{T}$ that would lead to an observable de-phasing at LISA, for the densities of the solid lines in Fig.~\ref{fig:maxdensity} (i.e., the maximum allowed by our perturbative treatment). Dashed lines:~$\mathcal{T}$ for~$\fR_{\rm A}$ and~$\fR_{\rm B}$, resp.,~Eqs.~\eqref{AQB_reflectivity} and~\eqref{QBH_reflectivity}. We use~$\upalpha=4 \ln 2$ as~\citet{Mukhanov:1986me} (see also~\cite{Bekenstein:1995ju, Bekenstein:1997bt}) and~$\psi=0$, with $\chi=0.9$.}
    \label{fig:reflective}
\end{figure}

We can also estimate the error of LISA in measuring $\delta \mathcal{T}$ (which carries information on the nature of the compact objects), through the de-phasing imparted by accretion on the inspiralling. Here, we do a very rough estimate using Eq.~\eqref{boundphi} and neglecting correlations of $\mathcal{T}$ with other parameters. For an EMRI (with $T_{\rm obs}=4\textrm{yr}$),
\begin{align}
\delta \mathcal{T}_2&\overset{(q\gg 1)}{\sim}10^{-4}\left(\frac{10^{-7}}{\rho_0 M_{\rm T}^2}\right) \left(\frac{M_{\rm T}}{10^6 M_\odot}\right)^{\frac{8}{3}} 
\nonumber
\\
&\qquad \times \left(\frac{f}{1\, {\rm mHz}} \right)^{\frac{8}{3}} \left(\frac{10^2}{{\rm SNR}}\right)~. 
\end{align}
The same estimate for the case of equal-mass supermassive BH binaries is shown in Fig.~\ref{fig:reflective}.
These results show that the observation in LISA of a loud EMRI (with SNR~$\gtrsim 10^2$) or a supermassive BH binary evolving in an ultra-light boson cloud can, in principle, be used to probe the nature of the compact objects involved.

If the primary of an EMRI sustains the cloud, one has~$\mu M_2 \sim 0.1/q$, resulting in~$\mathcal{T}_2\approx 0$ (i.e., behaving as a perfect mirror) for the reflectivities inspired by quantum corrections to classical BHs. Therefore, such EMRIs could distinguish a classical from a quantum BH, but could not be used to discriminate between different models of quantum BHs. Although supermassive BH binaries need larger~$\mathcal{T}$ as compared to EMRIs, they could discriminate between different models of quantum BHs (as shown in Fig.~\ref{fig:reflective}). Only area-quantized models with~$\upalpha \approx 4 \mu/T_{\rm H}$~($\lesssim 16$ for~$\chi \lesssim 0.9$) can be probed through environmental effects in the binary inspiral; these include~$\upalpha= 4 \ln 2$ proposed by~\citet{Mukhanov:1986me} (also~\cite{Bekenstein:1995ju, Bekenstein:1997bt}), but not~$\upalpha =8 \pi$~\cite{Bekenstein:1974jk, Maggiore:2007nq}.
Note that the observation of heavier binaries is not necessarily more effective in constraining~$\mathcal{T}$. Even though Fig.~\ref{fig:reflective} may suggest it, this is a consequence of our perturbative treatment being valid up to larger~$\rho M_{\rm T}^2$ for more massive binaries (see Fig.~\ref{fig:maxdensity}).

\section{Conclusions}%
In this paper, we derived analytical expressions for the accretion rate and DF [resp., Eqs.~\eqref{accret} and \eqref{drag_force}] acting on ECOs moving through ultra-light scalars. We used them then to estimate the orbital dephasing in the inspiralling of different binaries in the LISA band. Our simple estimate indicates that EMRIs and supermassive BH binaries may be used to probe quantum corrections to classical BHs. 
Other methods to probe the nature of ECOs in the inspiral stage include tidal deformability/heating~\cite{Maselli:2017cmm, Datta:2019epe, Sago:2021iku, Maggio:2021uge, datta2021imprint, Chakraborty:2022zlq, Nair:2022xfm, Chakraborty:2023zed, nair2023dynamical}. The observation of effects from tidal deformability is quite challenging, as they appear at high PN order ($\geq 5$ PN) and the Love numbers of ECOs are expected to be small~\cite{Maselli:2017cmm} (except if resonances are excited~\cite{Chakraborty:2023zed, Nair:2022xfm}). Tidal heating can be used to constrain the reflectivity to~$\delta \fR_1 \sim 10^{-4}$ for the primary of an EMRI similar to our fiducial one~\cite{Datta:2019epe, Chakraborty:2021gdf, Sago:2021iku, Maggio:2021uge}. The environmental effects studies in this work are sensitive instead to the secondary's reflectivity~$\fR_2$.

We have used various approximations which merit further research. Similarly to all the previous works on area-quantized BHs, we have neglected multi-particle processes (e.g.,~\cite{Reiss:1970}) in the BH excitation. However, the number of incident scalar particles per wavelength on the objects is of the order of $\sim \rho_0 M^2/(\hbar \mu^2) \approx 10^{70} (M/10 M_\odot)^{2}(\rho_0 M_{\rm T}^2/10^{-10})(0.1/\mu M_{\rm T})^2$. While we have neglected the backreaction on the environment, in certain regions of parameter space, this assumption may not be valid due to resonant depletion of the cloud~\cite{Baumann:2018vus, Baumann:2019ztm, Tomaselli:2024bdd, Boskovic:2024fga}. Nevertheless, scalar clouds can survive (at least) up to the LISA band in certain scenarios (see, e.g.,~\cite{Takahashi:2021eso, Guo:2024iye}), but very little is known about the cloud dynamics for symmetric binaries. When the backreaction is important, an interesting possibility is that the ECO reflectivity also impacts the resonant history of the binary, due to the broadening of the energy levels of the quasi-bound states~\cite{Zhou:2023sps}. In any case, our results apply to general scalar field configurations, like the global states that may form dynamically around massive BH binaries (see, e.g.,~\cite{Bamber:2022pbs, Aurrekoetxea:2023jwk}), or self-gravitating structures (like fuzzy dark matter solitons~\cite{Cardoso:2022nzc}).

\section*{Acknowledgements}

The authors are grateful to Vitor Cardoso for helpful discussions and useful comments on an earlier version of this work. S.C. acknowledges the warm hospitality at the Albert-Einstein Institute, where a part of this work was performed, which was supported by the Max-Planck-India mobility grant. Research of S.C. is supported by the MATRICS and the Core research grants from SERB, Government of India (Reg. Nos. MTR/2023/000049 and CRG/2023/000934). 
R.V. is supported by grant no. FJC2021-046551-I funded by MCIN/AEI/10.13039/501100011033 and by the European Union NextGenerationEU/PRTR. R.V. also acknowledges the support from the Departament de Recerca i Universitats from Generalitat de Catalunya to the Grup de Recerca `Grup de F\'isica Te\`orica UAB/IFAE' (Codi: 2021 SGR 00649).
J.C.F. acknowledges support from the R.O.C. (Taiwan) National Science and Technology Council (NSTC) grant No. 112-2811-M-002-132.

\appendix

\section{Energy and momentum in asymptotically flat spacetimes}\label{appnA}

In this section, we intend to provide some justification for the energy loss and force formulas employed in the main body of this article. A justification was originally provided in \cite{Clough:2021qlv}, however, that derivation is based on pseudotensorial quantities; here we briefly describe a tensorial generalization of that approach (based on the formalism developed in \cite{Lynden-Belletal1995} and \cite{Feng:2021lfa}) in which the meaning of the pseudotensorial quantities is elucidated. 

We begin by introducing a flat background metric $\bar{g}_{\mu\nu}$, and in this appendix, we employ the convention that indices are raised and lowered using $\bar{g}_{\mu\nu}$ and define $\hat{g}^{\mu \nu}$ to be the inverse of the dynamical metric $g_{\mu \nu}$ (we do not assume the usual Cartesian $\eta_{\cdot\cdot}=\text{diag}(-1,1,1,1)$ form, and permit curvilinear coordinates). We also introduce a compatible torsion-free reference connection $\bar{\nabla}$ with connection coefficients $\bar{\Gamma}{^\lambda}{_{\mu \nu}}$. One can construct the following tensorial quantities:
\begin{align} \label{WDefinition}
W{^\lambda}{_{\mu \nu}} &:= {\Gamma}{^\lambda}{_{\mu \nu}} 
                            - \bar{\Gamma}{^\lambda}{_{\mu \nu}}\\
W^\mu &:= \hat{g}^{\sigma \tau} W{^\mu}{_{\sigma \tau}}
            -
          \hat{g}^{\mu \sigma} W{^\nu}{_{\nu \sigma}}~,
\end{align}
which can be used to construct the following first-order action (cf. Eq. (28) in \cite{Feng:2021lfa}):
\begin{align} \label{GravitationalAction}
S_{GR} &= \frac{1}{2 \kappa}\int_{U} d\mu 
            \left[{R} - {\nabla}_\mu W^\mu\right] \\ \nn
        &= \frac{1}{2 \kappa}\int_{U} d\mu \left(W{^{\mu}}_{\sigma \lambda} W{^\lambda}{_{\mu \tau}} - W{^{\nu}}_{\sigma \tau} W{^\lambda}{_{\lambda \nu}}\right) \hat{g}^{\sigma \tau}~,
\end{align}
where $d\mu:=\sqrt{|g|}d\underline\mu$, $\kappa=8\pi G$, and the flatness of the background metric $\bar{g}_{\mu\nu}$ has been employed. Given a vector $\xi^\mu$ which is covariantly constant with respect to $\bar{g}_{\mu\nu}$ (namely, $\bar\nabla_\nu\xi^\mu=0$), the following current:
\begin{equation} \label{KOri}
    \begin{aligned}
    J^\mu
    = \, &
    \xi^\beta 
    \left[2 G_{\nu \beta} \, \hat{g}^{\mu \nu} - {_{\rm E}}\Theta{^\mu}{_\beta} \right]~,
    \end{aligned}
\end{equation}
identically satisfies $\partial_\mu(\sqrt{|g|} J^\mu)=0$ (by way of the Katz-Ori identity \cite{Lynden-Belletal1995}). The current can can be thought of as an energy density current for timelike $\xi^\mu$, and as a momentum density current for spacelike $\xi^\mu$, and one can replace the Einstein tensor using the Einstein equation $G_{\mu \nu}=\kappa T_{\mu \nu}$. The quantity ${_E}\Theta{^\alpha}{_\beta}$ is a tensorial generalization of the Einstein (canonical energy-momentum) pseudotensor:
\begin{align} \label{CEMT}
{_E}\Theta{^\alpha}{_\beta}=&
    - W^{\alpha} W{^\sigma}_{\sigma \beta}
    - \left[
        W{^\alpha}{_{\beta \sigma}} \hat{g}^{\sigma \tau} 
        + 
        W{^\tau}{_{\beta \sigma}} \hat{g}^{\alpha \sigma}
      \right]
    W{^\nu}_{\nu \tau} 
    \nonumber
    \\
    &+
    \left[ W^{\sigma}{_{\beta \nu}} \hat{g}^{\nu \tau} 
        + 
        W^{\tau}{_{\beta \nu}} \hat{g}^{\nu \sigma} \right]
    W^{\alpha}{_{\sigma \tau}}
    \nonumber
    \\
    &-
    \delta{^\alpha}{_\beta} 
    \left[{R} - {\nabla}_\sigma W^\sigma\right]~.
\end{align}
From Eq. \eqref{CEMT}, one can see that the ${_E}\Theta{^\alpha}{_\beta}$ shifts the pseudotensorial ambiguities of the Einstein pseudotensor to the choice of coordinates on the flat reference geometry used to construct $\bar{\Gamma}{^\lambda}{_{\mu \nu}}$. A detailed discussion of this point can be found in \cite{Feng:2021lfa}.

Now for convenience, one can define $\bar{J}^\mu:=\Psi J^\mu$, where $\Psi:=\sqrt{|g/\bar{g}|}$. Now what is particularly useful is the fact that the current $\bar{J}^\mu$ can be clearly split into a gravitational part and a nongravitational part $\bar{J}^\mu=\bar{J}_M^\mu + \bar{J}_G^\mu$ where:
\begin{equation} \label{Parts}
    \begin{aligned}
    \bar{J}_M^\mu
    =
    \, 
    2 \kappa \Psi
    \xi^\beta T_{\nu \beta} \, \hat{g}^{\mu \nu}~; \qquad
    \bar{J}_G^\mu
    =
    \,
    - 
    \Psi \xi^\beta 
    {_{\rm E}}\Theta{^\mu}{_\beta}~.
    \end{aligned}
\end{equation}
From the divergence theorem, one can derive the following expression for a divergence-free current satisfying $\bar{\nabla}_\mu \bar{J}^\mu=0$:
\begin{align} \label{ConservationLawsGen3}
    \partial_t\int_{\Sigma_{t}} d\bar{\Sigma}_\mu \bar{J}^\mu
    =
    -
    \int_{\partial \Sigma_{t}} d\bar{\sigma} r_\mu \bar{J}^\mu ~,
\end{align}
where $d\bar{\Sigma}_\mu$ is the directed surface element on $\Sigma_{t}$, and $d\bar{\sigma}$ is the surface element on $\partial \Sigma_{t}$ (both defined with respect to $\sqrt{|\bar{g}|}$), with $r^\mu$ is a unit vector tangent to $\Sigma_{t}$ and normal to $\partial\Sigma_{t}$. The derivation is straightforward, and follows that found in \cite{Clough:2021qlv}. One may rewrite Eq. \eqref{ConservationLawsGen3} as:
\begin{align} \label{ConservationLawsGen4}
    \partial_t\int_{\Sigma_{t}} d\bar{\Sigma}_\mu \bar{J}_M^\mu
    &+
    \partial_t\int_{\Sigma_{t}} d\bar{\Sigma}_\mu \bar{J}_G^\mu
    \nonumber
    \\
    &=
    -
    \int_{\partial \Sigma_{t}} d\bar{\sigma} r_\mu \bar{J}_M^\mu
    -
    \int_{\partial \Sigma_{t}} d\bar{\sigma} r_\mu \bar{J}_G^\mu  ~.
    \nonumber
\end{align}
We consider a steady state energy-momentum tensor, and coordinates in which $g_{\mu\nu}$ is stationary, and it follows that:
\begin{equation} \label{SteadyStateInt}
    \partial_t\int_{\Sigma_{t}} d\bar{\Sigma}_\mu \bar{J}_M^\mu
    =0~,
\end{equation}
and one has the expression:
\begin{align} \label{ConservationLawsGen4}
    \partial_t\int_{\Sigma_{t}} d\bar{\Sigma}_\mu \bar{J}_G^\mu
    = \,&
    -
    \int_{\partial \Sigma_{t}} d\bar{\sigma} r_\mu \bar{J}_M^\mu
    -
    \int_{\partial \Sigma_{t}} d\bar{\sigma} r_\mu \bar{J}_G^\mu  ~.
    \nonumber
\end{align}
The left hand side of the above can be interpreted as the time derivative of the integrated energy density for timelike $\xi^\mu$ and a force for spacelike $\xi^\mu$; in both cases, the evaluation consists of a surface integral over $\partial \Sigma_{t}$. 

In an asymptotically flat spacetime, one can evaluate the surface integral in the vicinity of spatial infinity $\partial \Sigma_{t}\rightarrow i_0$, treating deviations from flat spacetime (in some appropriate coordinate gauge) as perturbations. Assuming $T_{\mu \nu} \sim O(\epsilon)$ and $g_{\mu \nu}-\bar{g}_{\mu \nu} \sim {W^\sigma}_{\mu \nu} \sim O(\epsilon)$, one finds that to $O(\epsilon)$
\begin{equation} \label{PartsLO}
    \begin{aligned}
    \bar{J}_M^\mu
    \approx
    \,
    2 \kappa
    T{^\mu}{_\nu} \xi^\nu , \qquad\qquad
    \bar{J}_G^\mu
    \approx
    \, 
    0~,
    \end{aligned}
\end{equation}
and it follows that 
\begin{align} \label{ConservationLawsGen4}
    \partial_t\int_{\Sigma_{t}} d\bar{\Sigma}_\mu \bar{J}_G^\mu
    \approx \,&
    -
    2 \kappa \int_{\partial \Sigma_{t}} d\bar{\sigma} r_\mu  
    T{^\mu}{_\nu} \xi^\nu~.
\end{align}
Note that to $O(\epsilon)$, the quantity $W{^\lambda}_{\mu\nu}$ does not appear, and the only dependence on the background coordinate gauge arises through the surface element $d\bar{\sigma}$.

As discussed right after Eq. \eqref{KOri}, the current $J^\mu$ represents an energy density current for timelike $\xi^\mu$ and a momentum current for spacelike $\xi^\mu$. Given Eq. \eqref{SteadyStateInt}, and Eq. \eqref{ConservationLawsGen3}, one can use Eq. \eqref{ConservationLawsGen4} and the choice of coordinate basis vectors $\partial_t$ and $\partial_i$ to obtain the results:
\begin{align} \label{EqAppdx:Edot}
    \partial_t E
    \approx \,&
    -
    2 \kappa \int_{\partial \Sigma_{t}} d\bar{\sigma} r_\mu  
    T{^\mu}{_t}~,
\end{align}
and
\begin{align} \label{EqAppdx:Pdot}
    F^i:=
    \partial_t P^i
    \approx \,&
    -
    2 \kappa \int_{\partial \Sigma_{t}} d\bar{\sigma} r_\mu  
    T{^\mu}{_i}~,
\end{align}
both of which we employ in the main text.

\section{Scattering amplitudes of a scalar in the low-frequency regime}\label{appnB}

In this appendix, we will provide the explicit calculation of the scattering amplitude $(R^m_l/I^m_l)$ in low frequency approximation using matched asymptotic expansion technique, as mentioned in the main letter.
\subsection{Scalar perturbation in the near horizon region}

We will start by solving the radial perturbation equation in the near-horizon regime. For this purpose, it is useful to employ a change of the radial coordinate, 
\begin{equation}
\label{x}
x\equiv \frac{r-r_{+}}{r_{+} - r_{-}} \ll \frac{l+1}{\omega(r_{+}-r_{-})}~,
\end{equation}
which is dimensionless and the inequality simply follows from the fact that near the horizon $x\ll 1$ and in the small-frequency approximation $\omega(r_{+}-r_{-})\ll 1$, as well. In terms of this new dimensionless radial coordinate, the radial perturbation equation becomes, 
\begin{align}
\label{master_nearhorizon}
x(x+1)&\frac{d}{dx}\left[x(x+1)\frac{d}{dx}\mathcal{R}_{l}^{m~\textrm{(nh)}}\right] 
\nonumber
\\
&+\left[Q^2-l(l+1)x(x+1)\right]\mathcal{R}_{l}^{m~\textrm{(nh)}}=0~, 
\end{align}
where we have defined, 
\begin{align}\label{defQ}
Q&\equiv \frac{r_{+}^2 + a^2 }{r_{+}-r_{-}} \left(m \Omega_H - \omega\right)~,
\end{align}
and $\mathcal{R}_{l}^{m~\textrm{(nh)}}$ is the radial perturbation in the near-horizon region. 
The general solution of Eq. \eqref{master_nearhorizon} is given in terms of hypergeometric functions \cite{abramowitz1988handbook},
\begin{align}
\label{soln_nearhorizon}
&\mathcal{R}_{l}^{m~\textrm{(nh)}}=\left(1+x\right)^{iQ}
\nonumber
\\
&\times\bigg\{c_1 x^{-iQ}\,_{2}F_{1}(-l,l+1;1-2iQ;-x)
\nonumber
\\
&+c_2 x^{iQ}
\,_{2}F_{1}(-l+2iQ,l+1+2iQ;1+2iQ;-x)\bigg\}~.
\end{align}
The arbitrary constants $c_1$ and $c_2$ are to be fixed by taking the small $x$ limit of the above solution and matching it with the physical boundary condition at the horizon. In this limit, hypergeometric functions can be expanded to obtain a power law solution,
\begin{equation}
\label{nearhorizon_smallx}
\mathcal{R}_{l}^{m~\textrm{(nh)}}\Big|_{\rm near} = c_1 x^{-iQ} + c_2 x^{iQ}~.
\end{equation}
Noting that $\mathcal{R}_{l}^{m}=(r_{+}^{2}+a^{2})^{-1/2}\Psi_{l}^{m}$, in the near horizon limit, as well as the fact that $Q\ln x=-(\omega-m\Omega_{\rm h})r_{*}$, we obtain,
\begin{align}
\label{def_c1c2}
c_1=\frac{\mathbb{R}_{l}^{m}}{\sqrt{r_{+}^2 +a^2}}~;
\qquad
c_2=\frac{T_{l}^{m}}{\sqrt{r_{+}^2 +a^2}}~.
\end{align}
The BH scenario can be obtained from our results by simply substituting the surface reflectivity $\mathbb{R}_{l}^{m}$ of the ECO to be zero, implying $c_1=0$. In what follows we will keep the results as general as possible and are not going to assume any particular form of the reflectivity, by keeping both $c_1$ and $c_2$ explicitly in our calculations.

Having discussed the near-horizon behaviour of the near-horizon solution, we consider now the extension of the near-horizon solution to the intermediate region, which is obtained by taking the $x\gg 1$ limit of Eq. \eqref{soln_nearhorizon}. Therefore, expanding the hypergeometric functions for large values of $x$ and keeping only the leading order terms, we obtain~\cite{Starobinsky:1973aij, Dey:2020lhq},
\begin{align} 
\label{nearhorizon_large_x}
\mathcal{R}_{l}^{m~\textrm{(nh)}}\Big|_{\rm far}&\simeq d_{1}x^l+\left(\frac{d_{2}}{2l+1}\right)x^{-l-1}~.
\end{align}
\begin{widetext}
\noindent
Here we have denoted the coefficients of the terms $x^l$ and $x^{-l-1}$ are  by $d_1$ and $d_2$ and these can be expressed in terms of $c_1$ and $c_2$ as, 
\begin{align}
d_1&=\frac{\Gamma(2l+1)}{\Gamma(l+1)} \left[c_1\frac{\Gamma(1-2iQ)}{\Gamma(l+1-2iQ)} 
+c_2\frac{\Gamma(1+2iQ)}{\Gamma(l+1+2iQ)}\right]~,
\\
d_2&=(2l+1)\frac{\Gamma(-2l-1)}{\Gamma(-l)}\left[c_1\frac{\Gamma(1-2iQ)}{\Gamma(-l-2iQ)} 
+c_2\frac{\Gamma(1+2iQ)}{\Gamma(-l+2iQ)}\right]~.
\end{align}
Using the explicit forms of $c_1$ and $c_2$ from Eq. \eqref{def_c1c2} along with various identities of gamma functions, one can express $d_1$ and $d_2$ in terms of the reflection and transmission coefficients associated with the surface of the ECO as,
\begin{align}
d_1&=\frac{(2l)!}{l!}\frac{1}{\sqrt{r_{+}^2 +a^2}}\left[\frac{\mathbb{R}_{l}^{m}}{(1-2iQ)_l}+\frac{T_{l}^{m}}{(1+2iQ)_{l}}\right]~,
\label{d1}
\\
d_2&=\frac{(-1)^{l+1}}{2}\frac{l!}{(2l)!}\frac{1}{\sqrt{r_{+}^2 +a^2}}
\Big[\mathbb{R}_{l}^{m}(-l-2iQ)_{l+1}+T_{l}^{m}(2iQ-l)_{l+1}\Big]~,
\label{d2}
\end{align}
where, we have used the Pochhammer symbol, defined as, $(x)_n=x(x+1)(x+2)\cdots(x+n-1)$. Further, note that we also have the following result: $(-l-2iQ)_{l+1}=(-l-2iQ)\cdots(-2iQ)=(-1)^{l+1}(2iQ)_{l+1}$, and hence the above coefficients can also be expressed in various other forms. Further, in the BH limit ($\mathbb{R}_{l}^{m}\to 0$), the above solution reduces to the one in \cite{vicente2022dynamical}. 
\end{widetext}
In the intermediate region, with $x \gg 1$, we can effectively express the re-scaled radial coordinate as follows, $x\simeq\{r/(r_{+}-r_{-}\}$, such that we can rewrite the near-horizon solution in the matching region as,
\begin{align}
\label{nearhorizon_matching_r}
\mathcal{R}_{l}^{m~\textrm{(nh)}}\Big|_{\rm far} &\approx \left(\frac{d_1}{(r_{+} - r_{-})^l}\right)r^{l}
\nonumber
\\
&+\left(\frac{d_2}{(r_{+} -r_{-})^{-l-1}}\right)r^{-l-1}\,.
\end{align}
We will be using this result to match with the far zone solution, which we will now derive.  

\subsection{Scalar perturbation in the asymptotic region}

In this section, we will present the solution to the radial part of the scalar perturbation in the asymptotic region, which corresponds to $r \gg r_{+}$. In this limit, the radial perturbation equation simplifies to, 
\begin{equation} 
\label{master_asymp}
\left[\frac{d^2}{d r^2}+k_\infty^2-\frac{2\eta k_\infty}{r}-\frac{l(l+1)}{r^2}\right]\left(\sqrt{\Delta}\, \mathcal{R}_{l}^{m~\textrm{(asy)}}\right)=0~.
\end{equation}
The general solutions to the above equation is given in terms of Coulomb wave functions \cite{NIST:DLMF},
\begin{equation} 
\label{soln_asymp}
\mathcal{R}_{l}^{m~\textrm{(asy)}}=\frac{c_3}{r}\,\text{F}_l^C(\eta, k_\infty r)
+\frac{c_4}{r}\,\text{G}_l^C(\eta, k_\infty r)~.
\end{equation}
Now we need to match this asymptotic solution with the near-horizon solution in the intermediate zone. For this purpose, we expand the Coulomb functions in the limit $k_\infty r \ll l$ (which corresponds to the radial parameter being small compared to the impact parameter), and one recovers,
\begin{align} 
\label{soln_asymp_matching}
\mathcal{R}_{l}^{m~\textrm{(asy)}}\Big|_{\rm near}&\simeq c_3 C_l(\eta)k_\infty^{l+1}\,r^l
\nonumber
\\
&+c_4\left(\frac{k_\infty^{-l}}{(2l+1)C_l(\eta)}\right)\,r^{-l-1}~,
\end{align}
where we have introduced the constant $C_{l}(\eta)$, having the following expression,
\begin{equation}
\label{def_C}
C_{l}(\eta)=\frac{2^l e^{-\eta \pi/2} |\Gamma(l+1+i \eta)|}{(2l+1)!}~.
\end{equation}
The large $r$ limit of the near-horizon solution, presented in Eq. \eqref{nearhorizon_matching_r}, and the small $r$ limit of the asymptotic solution, as discussed in Eq. \eqref{soln_asymp_matching}, must be matched to one another. This relates the arbitrary constants $c_{3}$ and $c_{4}$ to $d_{1}$ and $d_{2}$, such that,
\begin{align}
c_3&=\left\{d_1/C_l(\eta)\right\}k_\infty^{-l-1} (r_{+}-r_{-})^{-l}~,
\label{c_3}
\\
c_4&=d_2 C_l(\eta) k_\infty^{l} (r_{+}-r_{-})^{l+1}~.
\label{c_4}
\end{align}
Thus, we have related the near-horizon boundary condition to the asymptotic solution, and finally, to obtain the scattering amplitude $(R_{l}^{m}/I_{l}^{m})$ in terms of properties in the near-horizon regime, we must relate the coefficients $c_3$ and $c_4$ with the physical boundary condition at infinity. For this purpose, we need to consider the $k_\infty r \to \infty$ limit of the asymptotic solution in Eq. \eqref{soln_asymp}, which gives,
\begin{align}
\label{infinity_asymtot}
\mathcal{R}_{l}^{m~\textrm{(asy)}}\Big|_{\rm far}&\simeq \frac{c_3}{r} \sin [\theta_l(\eta, k_\infty r)]
\nonumber
\\
&+\frac{c_4}{r} \cos [\theta_l(\eta, k_\infty r)]\,.
\end{align}
Here, $\theta_l\equiv k_\infty r- \eta \log(2 k_\infty r)-l(\pi/2)+ \arg \Gamma(l+1+i \eta)$, where, $\arg(z)$ denotes the principal argument of the complex quantity $z$. Expanding out the $\sin$ and $\cos$ in terms of exponentials and using the asymptotic boundary condition, we obtain, 
\begin{align}
\label{B23}
R_{l}^{m}=\left(\frac{c_4 - ic_3}{2}\right)e^{-iA}~,
I_{l}^{m}=\left(\frac{c_4 + ic_3}{2}\right)e^{iA}\,.
\end{align}
where, we have defined $A\equiv (l\pi/2)-\arg \left[\Gamma(l+1+i\eta)\right]$ with $l$ taking integer values. 

\subsection{Matching region and scattering amplitudes}

Having derived all the relevant limits of the near-horizon and the asymptotic solutions, we wish to determine the scattering amplitudes by matching them appropriately. Note that the asymptotic amplitudes are dependent on the constants $c_{3}$ and $c_{4}$, which are dependent on the near-horizon quantities $d_{1}$ and $d_{2}$ through Eq. \eqref{c_4}, and to the horizon reflectivity by Eq. \eqref{d1} and Eq. \eqref{d2}, respectively. To obtain the scattering amplitude, we start by noting, $e^{-il \pi}=(-1)^l$, for integer values of $l$, which lets us obtain,
\begin{equation}
\label{exponentA}
e^{-2iA}=(-1)^l e^{2i\arg \Gamma(l +1 + i\eta)}~.
\end{equation}
Using this result for $e^{-2iA}$, along with Eq. \eqref{c_4} and Eq. \eqref{B23}, we obtain the following ratio between the scattering amplitudes, for generic choices of the angular momentum $l$,
\begin{align}
\label{RbyIApp}
\frac{R_{l}^{m}}{I_{l}^{m}}&=(-1)^l e^{2 i \arg \Gamma(l +1 + i\eta)} \left( \frac{c_4- i c_3}{c_4 + i c_3} \right)
\nonumber
\\
&= (-1)^l e^{2 i \arg \Gamma(l +1 + i\eta)}\left(\frac{z_l \frac{d_2}{d_1} - i}{z_l \frac{d_2}{d_1} + i} \right)
\nonumber
\\ 
&=(-1)^{l+1} e^{2 i \arg \Gamma (1+l+ i \eta)}
\nonumber
\\
\nonumber
\\
&\hskip -1 cm\times 
\left[\frac{(1+\sqrt{\mathfrak{R}_{l}^{m}})+\left(\frac{(l!)}{(2l)!}\right)^2 z_l Q |(1+ \bar{Q})_l|^2 (1-\sqrt{\mathfrak{R}_{l}^{m}})}{(1+\sqrt{\mathfrak{R}_{l}^{m}}) - \left(\frac{(l!)}{(2l)!}\right)^2 z_l Q |(1+ \bar{Q})_l|^2 (1-\sqrt{\mathfrak{R}_{l}^{m}})}\right]~.
\end{align}
Note that in the above expression we have defined the effective reflectivity $\sqrt{\mathfrak{R}_{l}^{m}}$ as the multiplication of the ratio $(\mathbb{R}_{l}^{m}/T_{l}^{m})$ with the ratio of Pochhammer symbols $\{(1+2iQ)_{l}/(1-2iQ)_{l}\}$, such that,
\begin{align}\label{def_reflect}
\sqrt{\mathfrak{R}_{l}^{m}}\equiv \frac{\left(1+2iQ\right)_{l}}{\left(1-2iQ\right)_{l}}\frac{\mathbb{R}_{l}^{m}}{T_{l}^{m}}~.
\end{align}
Moreover, we have used the following expression for the quantity $z_l$, 
\begin{equation}
\label{z_l-def}
z_{l}=C_{l}(\eta)^{2}k_{\infty}^{2l+1}(r_{+}-r_{-})^{2l+1}~,
\end{equation}
along with  $(r_{+}-r_{-})=2\sqrt{M^{2}-a^{2}}$. Putting the explicit expressions and after some manipulation, one can finally obtain the scattering amplitude,
\begin{widetext}
\begin{align}
\label{RbyI}
\frac{R_{l}^{m}}{I_{l}^{m}}&=(-1)^{l+1}e^{2i\arg \Gamma(1+l+ i \eta)}
\Bigg[\frac{1+\frac{(l!)^{2}}{(2l!)^{2}}QC_{l}(\eta)^{2}|(1+2iQ)_{l}|^{2}\left(2k_{\infty}\sqrt{M^{2}-a^{2}}\right)^{2l+1}\left(\frac{1-\sqrt{\mathfrak{R}_{l}^{m}}}{1+\sqrt{\mathfrak{R}_{l}^{m}}}\right)}{1-\frac{(l!)^{2}}{(2l!)^{2}}QC_{l}(\eta)^{2}|(1+2iQ)_{l}|^{2}\left(2k_{\infty}\sqrt{M^{2}-a^{2}}\right)^{2l+1}\left(\frac{1-\sqrt{\mathfrak{R}_{l}^{m}}}{1+\sqrt{\mathfrak{R}_{l}^{m}}}\right)}\Bigg]\,.
\end{align}
As evident, for vanishing $\mathbb{R}_{l}^{m}$, the effective reflectivity $\sqrt{\mathfrak{R}^{m}_{l}}$ also vanishes. Note that, we have not so far used the small-frequency approximation, which we wish to impose now. It turns out that the above approximation is easiest to apply for static situations, i.e., for compact objects with zero rotation. In this case, $Q=-2M\omega\ll1$, and hence we obtain, $|(1+2iQ)_{l}|=(l!)^{2}+\mathcal{O}(\omega^{2})$. Substituting all of these results in the above ratio of scattering amplitude, we obtain, 
\begin{align}
\label{RbyIstatic}
\left(\frac{R_{l}^{m}}{I_{l}^{m}}\right)_{\rm static}\simeq(-1)^{l+1}e^{2i\arg \Gamma(1+l+ i \eta)}
\Bigg[1-2\frac{(l!)^{4}}{(2l!)^{2}}C_{l}(\eta)^{2}\left(2k_{\infty}M\right)^{2l+2}\left(\frac{1-\sqrt{\mathfrak{R}_{l}^{m}}}{1+\sqrt{\mathfrak{R}_{l}^{m}}}\right)\left(\frac{\omega}{k_{\infty}}\right)\Bigg]\,,
\end{align}
where we have ignored terms $\mathcal{O}(\omega^{2})$. Moreover, in the above calculation, we have used the following approximations: $\{(1+x)/(1-x)\}\approx 1+2x$, along with the result that $Q=-2M\omega$. Further, the effective reflectivity $\sqrt{\mathfrak{R}^{l}_{m}}$ can be expressed as, 
\begin{align}
\label{reflec_expansion}
\sqrt{\mathfrak{R}^{l}_{m}}=\frac{\mathbb{R}_{l}^{m}}{T_{l}^{m}}\left[\frac{1-4iM\omega\frac{l(l+1)}{2(l!)}}{1+4iM\omega\frac{l(l+1)}{2(l!)}}\right]
\simeq \frac{\mathbb{R}_{l}^{m}}{T_{l}^{m}}\left[1-4iM\omega\frac{l(l+1)}{(l!)}+\mathcal{O}(\omega^{2})\right]~.
\end{align}
Thus, to the leading order in the frequency, the effective reflectivity $\mathfrak{R}_{l}^{m}$ is indeed the reflectivity of the surface of the ECO. As expected, the $\mathbb{R}_{l}^{m}\to 0$, implies $\mathfrak{R}_{l}^{m}\to 0$, and thus we recover the scattering amplitudes presented in \cite{vicente2022dynamical}. 

On the other hand, for rotating ECO, the quantity $Q$ defined in Eq. \eqref{defQ} depends on the rotation parameter and is not necessarily small. Thus, unlike the static case, we can no longer assume $Q\ll 1$, rather we have to content ourselves with the small frequency approximation alone. Therefore, we need to expand the ratio of scattering amplitudes in powers of $k_{\infty}M$, yielding,
\begin{align}
\label{RbyIrot}
\frac{R_{l}^{m}}{I_{l}^{m}}&=(-1)^{l+1}e^{2i\arg \Gamma(1+l+ i \eta)}
\Bigg[1+\frac{2(l!)^{2}}{(2l!)^{2}}QC_{l}(\eta)^{2}|(1+2iQ)_{l}|^{2}\left(2k_{\infty}\sqrt{M^{2}-a^{2}}\right)^{2l+1}\left(\frac{1-\sqrt{\mathfrak{R}_{l}^{m}}}{1+\sqrt{\mathfrak{R}_{l}^{m}}}\right)\Bigg]\,.
\end{align}
From the above expression it is easy to see that $l \neq 0$ modes will always contribute higher powers of $\omega M$, through the term $|(1+2iQ)_{l}|^{2}\left(2k_{\infty}\sqrt{M^{2}-a^{2}}\right)^{2l+1}$. Similarly, the expansion of reflectivity in Eq. \eqref{reflec_expansion} also explicitly suggests that higher $l$ modes contributes to higher powers of $\mathcal{O}(\omega M)$. Combined these two factor together, it follows that in the low-frequency regime, contribution of the $l=0$ mode will be dominant. We have used this result in the main letter as well. Considering only the contribution of $l = 0$ mode then, we rewrite the scattering amplitude as 
\begin{align}
\label{Rbyl0}
\frac{R^{0}_{0}}{I^{0}_{0}}&=-e^{2i\arg \Gamma(1+i\eta)}
\Bigg[1+4QC_{0}(\eta)^{2}k_{\infty}\sqrt{M^{2}-a^{2}}\left(\frac{1-\sqrt{\mathfrak{R}^{0}_{0}}}{1+\sqrt{\mathfrak{R}^{0}_{0}}}\right)\Bigg]\,,
\end{align}
where $\sqrt{\mathfrak{R}^{0}_{0}}=(\mathbb{R}^{0}_{0}/T^{0}_{0})$. This is the final result, which along with the expressions of $Q$ and $C_l(\eta)$ given in Eq. \eqref{defQ} and Eq. \eqref{def_C} respectively, have been used in main text. 

\section{Energy flux and Force experienced in the rest frame of the compact object: Low-frequency limit}\label{appnC}

In this appendix, we will explicitly determine the energy flux and force defined in the main text, experienced by the compact object in its rest frame, due to scattering of the scalar field with the ECO. Let us start with the discussion of energy flux. One can use the stress-energy tensor of scalar field,
\begin{equation}
\label{SET_field}
T_{\alpha\beta} = \nabla_\alpha \Phi^* \nabla_\beta \Phi - \frac{1}{2} g_{\alpha\beta} \left( \nabla_\delta \Phi^* \nabla^\delta \Phi + \mu^2 \Phi  \right)~,
\end{equation}
which yields the following expression for the rate of energy absorption, 
\begin{align} 
\label{dotE_1}
\dot{E}&=\frac{\pi \hbar\omega n}{\mu k_\infty}\sum_{l,m}(2l+1)\frac{(l-m)!}{(l+m)!}\bigg(1-\bigg|\frac{R_{l}^{m}}{I_{l}^{m}}\bigg|^2\bigg) \times \Big|{P_S}_l^m(\cos\vartheta,\gamma^2)\Big|^2~.
\end{align}
In the low-frequency limit $l=0=m$ mode will contribute the most, so one can obtain the energy flux expression used in main text by using Eq. \eqref{Rbyl0} in the above equation for $l=0=m$ mode. 

Next let us look at the expression for imparted force on a compact object due to \df. Starting from the expression of the force experienced by the BH, as elaborated in the main text, and using the stress-energy tensor of the scalar field Eq. \eqref{SET_field} along with the boundary condition at spatial infinity yields,
\begin{align}
F^x&\simeq 0~,
\\
F^y&\simeq  \frac{4\pi \hbar n}{2 \mu}\sum_{l, m} \frac{(l-m)!}{(l+m)!} P_l^{m}(\cos \vartheta) P_{l+1}^{m+1}(\cos \vartheta) \times \Re \left[ 1+ \left( \frac{R^{m}_l}{I^{m}_l} \right)^* \frac{R^{m}_{l+1}}{I^{m}_{l+1}}\right]~,
\\
F^z &\simeq -\frac{4\pi \hbar n}{2\mu} \sum_{l, m} \frac{(l-m+1)!}{(l+m)!} P_l^m(\cos \vartheta) P_{l+1}^m(\cos \vartheta) \times \Re \left[ 1+ \left( \frac{R^{m}_l}{I^{m}_l} \right)^* \frac{R^{m}_{l+1}}{I^{m}_{l+1}} \right]~.
\end{align}
For convenience, it is customary to transform our Cartesian frame of reference so that the scalar wave is incident on the \co\, along $\bm{\xi} = -\partial_{z'}$, where the prime denotes the transformed coordinate system. To take the asymptotically Cartesian coordinate system $(x,y,z)$ to the rotated coordinate system $(\partial_{x'},\partial_{y'},\partial_{z'})$, we consider a rotation around $\partial_x$ by an angle $\vartheta$. Such that the force acting on the \co\ in the primed coordinate system can then be written as,
\begin{align}
F^{x'}&=F^x = 0~,
\label{rotatedFx}
\\
F^{y'}&=\cos\vartheta F^y + \sin \vartheta F^z = 0~,
\label{rotatedFy}
\\
F^{z'}&=\cos \vartheta F^z - \sin \vartheta F^y =-\sum_{l,m} \frac{2 \pi \hbar n}{\mu} (l+1)\,\textrm{Re}\left[ 1+ \left( \frac{R^{m}_l}{I^{m}_l} \right)^* \frac{R^{m}_{l+1}}{I^{m}_{l+1}} \right].
\label{rotatedFz}
\end{align}
In arriving at the above expressions, we have used various identities involving Legendre polynomials, which can be found in \cite{vicente2022dynamical}. Using the ratio of scattering amplitudes from Eq. \eqref{RbyIrot}, we obtain,
\begin{align}
&\textrm{Re}\left[1+\left(\frac{R_l^{m}}{I_l^{m}}\right)^{*}\frac{R_{l+1}^{m}}{I_{l+1}^{m}}\right]
=\textrm{Re}\Bigg\{1-e^{2i\left[\arg \Gamma(2+l+i\eta)-\arg \Gamma(1+l+i\eta)\right]}
\nn
\\
&\qquad \times \Bigg[1+\frac{2(l!)^{2}}{(2l!)^{2}}QC_{l}(\eta)^{2}|(1+2iQ)_{l}|^{2}\left(2k_{\infty}\sqrt{M^{2}-a^{2}}\right)^{2l+1}\left(\frac{1-\sqrt{\mathfrak{R}^{l}_{m}}}{1+\sqrt{\mathfrak{R}^{l}_{m}}}\right)^{*}\Bigg]
\nn
\\
&\qquad \times \Bigg[1+\frac{2[(l+1)!]^{2}}{[2(l+1)!]^{2}}QC_{l+1}(\eta)^{2}|(1+2iQ)_{l+1}|^{2}\left(2k_{\infty}\sqrt{M^{2}-a^{2}}\right)^{2l+3}\left(\frac{1-\sqrt{\mathfrak{R}^{l+1}_{m}}}{1+\sqrt{\mathfrak{R}^{l+1}_{m}}}\right)\Bigg]\Bigg\}\,.
\end{align}
Here the reflectivity $\sqrt{\mathfrak{R}^{l}_{m}}$ depends on the azimuthal number $l$, in general, which holds true for the models of quantum BH discussed in the main text. Moreover, in the low-frequency limit, the most dominant contribution comes from the lowest-lying $l=0$ mode, and we have the following result: $\arg \Gamma(2+l+i\eta)-\arg \Gamma(1+l+i\eta)=\arg [(1+l+i\eta)\Gamma(1+l+i\eta)]-\arg \Gamma(1+l+i\eta)=\arg (1+l+i\eta)$, where we have used the identity: $\arg (z_{1}z_{2})=\arg z_{1}+\arg z_{2}$. Using these, the above expression simplifies to,
\begin{align}\label{force1higher}
&\textrm{Re}\left[1+\left(\frac{R_l^{m}}{I_l^{m}}\right)^{*}\frac{R_{l+1}^{m}}{I_{l+1}^{m}}\right]
=\textrm{Re}\Biggl\{1-e^{2i\arg (1+l+i\eta)}
\Bigg[1+4QC_{0}(\eta)^{2}k_{\infty}\sqrt{M^{2}-a^{2}}\left(\frac{1-\sqrt{\mathfrak{R}^{0}_{0}}}{1+\sqrt{\mathfrak{R}^{0}_{0}}}\right)^{*}\Bigg]\biggr\}\,. 
\end{align}
Here, we have used the low-frequency approximation and have ignored all the higher-order terms in $M\omega$, such that only the $l=0$ mode contributes. To simplify further, we notice that $l=0$ implies $m=0$, such that $Q\propto \omega$, and expressing $\alpha_{l}=2\arg (1+l+i\eta)$, Eq. \eqref{force1higher} can be rewritten as,
\begin{align}\label{force11}
\textrm{Re}\Bigg[1&+\left(\frac{R_l^{m}}{I_l^{m}}\right)^{*}\frac{R_{l+1}^{m}}{I_{l+1}^{m}}\Bigg]
\approx \textrm{Re}\Biggl\{1-\left(\cos\alpha_{l}+i\sin \alpha_{l}\right) 
\Bigg[1-\textrm{Re}\left(\frac{1-\sqrt{\fR^{0}_{0}}}{1+\sqrt{\fR^{0}_{0}}}\right)\frac{\omega k_{\infty}A_{+}}{2\pi}\left(\frac{\pi \eta e^{-\pi \eta}}{\sinh(\pi \eta)}\right)\Bigg]\Biggr\}
\nn
\\
&=2\sin^{2}\left(\frac{\alpha_{l}}{2}\right)+\textrm{Re}\left(\frac{1-\sqrt{\fR^{0}_{0}}}{1+\sqrt{\fR^{0}_{0}}}\right)\frac{\omega k_{\infty}A_{+}}{2\pi}\left(\frac{\pi \eta e^{-\pi \eta}}{\sinh(\pi \eta)}\right)~,
\end{align}
where, in the terms depending on $\omega$, we have expanded the $\sin \alpha_{0}\approx \alpha_{0}$ and $\cos \alpha_{0}\approx 1-(\alpha_{0}^{2}/2)$. Note that the angle $\alpha_{0}=2\tan^{-1}\eta\sim 2\eta \ll 1$. Given the above expression for the ratio of scattering amplitudes, we can express the force, as presented in Eq. \eqref{rotatedFz}, on the \co\ due to \df\ experienced in the DM environment as,
\begin{align}\label{force22}
F^{z'(\textrm{low})}_{\rm CO}&=-\frac{4\pi\hbar n}{\mu}\sum_{l}(l+1)\sin^{2}\left(\frac{\alpha_{l}}{2}\right)
-\frac{4\pi\hbar n}{\mu}\left(\frac{\pi \eta e^{-\pi \eta}}{\sinh(\pi \eta)}\right)\frac{\omega k_{\infty}A_{+}}{4\pi}\textrm{Re}\left(\frac{1-\sqrt{\fR^{0}_{0}}}{1+\sqrt{\fR^{0}_{0}}}\right)
\nn
\\
&=-\frac{4\pi\hbar n}{\mu}\Biggr\{\sum_{l\geq 1} \frac{\eta^2 l}{\eta^2 + l^2} +\frac{\omega k_\infty A_{+}}{4 \pi}  
\left(\frac{\pi \eta e^{- \pi \eta}}{\sinh (\pi\eta)}\right)\textrm{Re}\left(\frac{1-\sqrt{\fR^{0}_{0}}}{1+\sqrt{\fR^{0}_{0}}}\right)\Biggr\}~.
\end{align}
\end{widetext}
Note that we have used the identity: $\sin(\alpha_{l-1}/2)=\{\eta/\sqrt{l^{2}+\eta^{2}}\}$ in arriving at the last line of the calculation, presented above. The first term dependent on the azimuthal number $l$ leads to a logarithmic divergence with $l$ in the expression for the force, due to the summation involved, and hence we introduce a cutoff $l_{\textrm{max}}$ which is related to the impact parameter $b_{\text{max}}$ of the scalar wave through the following relation,
\begin{equation}\label{max_impact}
b_{\text{max}}=\frac{\sqrt{l_{\text{max}}(l_{\text{max}} +1)}}{k_\infty}\,.
\end{equation}
As a consequence, the sum over all azimuthal numbers $l$ must be truncated and the resulting truncated sum can be written in terms of the Digamma function $\psi(z)$. This is the result we have used in the main text, keeping only the leading order terms in frequency. 

\bibliography{biblio1}

\newpage

\end{document}